\documentclass[useAMS,usenatbib,usegraphicx]{mn2e}
\usepackage{amssymb}
\usepackage{graphicx}
\usepackage{subfigure}
\usepackage{longtable}

\newif\ifAMStwofonts

\def\kms{km\,s$^{-1}$}

\newcommand{\hi}{H\,{\sc i}}
\newcommand\hii{H\,{\sc ii}~}

\newcommand\etal{et al.~}

\title{Sh2-205: II. Its quiescent stellar formation activity}
\author[G.A. Romero and C.E. Cappa]
       {G.A. Romero$^{1,2}$\thanks{E-mail: gisela@dfa.uv.cl, Post-doc fellow of Universidad de Valpara\'{\i}so, Chile} and C.E. Cappa$^{2}$\thanks{Member of the Carrera del Investigador Cient\'{\i}fico of CONICET, Argentina} \\
      $^{1}$ Departamento de F\'{\i}sica y Astronom\'{\i}a, Universidad de Valpara\'{\i}so, Avenida Gran Breta\~na
      1111, Valpara\'{\i}so, Chile\\
      $^{2}$Instituto Argentino de Radioastronom\'{\i}a, CCT-La Plata, CONICET, CC 5, 1894 Villa Elisa,
        Argentina,                                             \\
      Facultad de Ciencias Astron\'omicas y Geof\'{\i}sicas,
       Universidad Nacional de La Plata,                      
       Paseo del Bosque s/n, 1900 La Plata, Argentina \\
      }
   
\pagerange{\pageref{firstpage}--\pageref{lastpage}}

\pubyear{2008}

\begin{document}

\maketitle

\label{firstpage}

\begin{abstract}
We present a study of active stellar forming regions in the environs of the \hii region Sh2-205.
The analysis is based on data obtained from point source catalogues and images extracted from 2MASS, MSX, and IRAS surveys. Complementary data are taken from CO survey. The identification of primary candidates to stellar formation activity is made following colour criteria and the correlation with molecular gas emission.

A number of stellar formation tracer candidates are projected on two substructures of the \hii region: SH\,148.83--0.67 and SH\,149.25--0.00. However, the lack of molecular gas related to these structures casts doubts on the nature of the sources. Additional infrared sources may be associated with the \hi~shell centered at  \hbox{{\it (l,b)} = (149\degr 0\arcmin, --1\degr 30\arcmin)}. 

The most striking active area was found in connection to the \hii region \hbox{LBN 148.11--0.45}, where stellar formation candidates are projected onto molecular gas. The analytical model to the "collect and collapse" process shows that stellar formation activity could have been triggered by the expansion of this \hii region.
\end{abstract}


\section{Introduction}\label{intro}

\begin{figure}
\centering
\includegraphics[angle=0,width=0.4\textwidth]{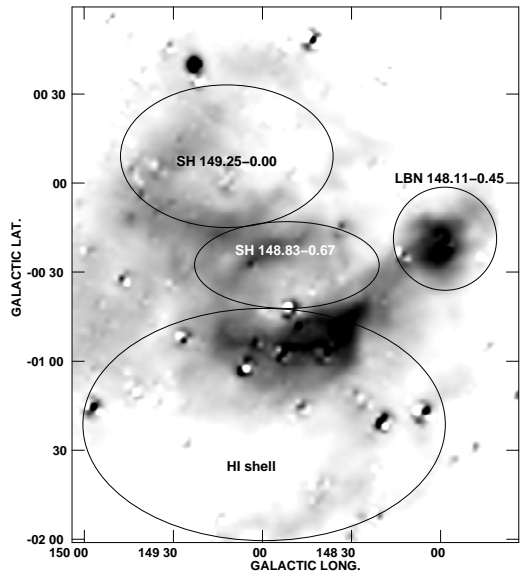}
 \caption{\small{Overview of Sh2-205 region: the areas discussed in the paper are schematically indicated in the VTSS continuun image.}}
 \end{figure}

Observational studies have shown that expanding \hii regions and interstellar bubbles are surrounded by massive and dense slowly expanding shells. Stellar formation may be favoured within these dense envelopes around
\hii regions (Deharveng, Zavagno \& Caplan 2005) and stellar wind bubbles through the {\it collect
and collapse} mechanism first described by Elmegreen \& Lada (1977), which was analytically developed by Whitworth \etal (1994). Numerical simulations were performed by Dale, Bonnell \& Whitworth (2007). In \hii regions, this process can be summarized as follows: The supersonic expansion of an \hii region on their surroundings creates compressed layers where gas and dust are piled up between the ionization and the shock fronts. This shocked material grows in mass and may become gravitationally unstable. Under these conditions, the dense envelopes may break up forming massive cores which could become in nurseries for new generations of massive stars. Thus, the dynamical evolution of \hii regions provides the conditions for the star formation process to develop (Deharveng \etal2003, Zavagno \etal2006).
                                       
A similar process may act in the dense envelopes around interstellar bubbles. Indeed, young stellar object (YSO) candidates  have been found projected onto the neutral shells associated with some wind bubbles (e.g. Cappa \etal2005).
                                                                     
In a previous paper, we studied the ISM in the environs of Sh2-205, located at a distance of $\sim$ 1.0 kpc (Romero \& Cappa~2008, hereinafter referred to as Paper I). In that work, we showed the presence of three independent optical nebulae: \hbox{SH\,149.25--0.00}, \hbox{SH\,148.83--0.67}, \hbox{LBN\,148.11--0.45}, and an \hi~shell centered at \hbox{{\it (l,b)} = (149\degr 0\arcmin, --1\degr 30\arcmin)}.
For the sake of clarity, the areas discussed in the paper are indicated in Fig. 1. We determined that SH\,148.83--0.67 is an interstellar bubble powered by the stellar winds of HD\,24431. The origin of SH\,149.25--0.00, which can hardly be distinguished by its optical and faint radio emissions, remains an open question since no stellar object was found associated with the nebula. The shell centered at \hbox{{\it (l,b)} = (149\degr 0\arcmin, --1\degr 30\arcmin)}, of $\approx$ 2\fdg2 $\times$ 1\fdg5 in size, is placed at a kinematical distance of 1.5 kpc and may be related to the open cluster NGC\,1444.

The most striking area is LBN\,148.11--0.45. This is a classical \hii region of 30\arcmin $\times$ 24\arcmin~in size and \hbox{4 $\times$ 10$^{6}$ yr} in age. Neutral atomic and molecular gas in the velocity range \hbox{[--0.65, --11.1]} \kms~partially encircles this ionized region. It is also closely surrounded by a dust ring detected in the mid- and far-infrared, suggesting the presence of a PDR (photodissociation region) at the interface between the ionized and molecular regions. The derived ambient density (n$_{o}$ $\approx$ 800 cm$^{-3}$) indicates that this \hii region is evolving in a dense interstellar medium (Paper I).

The morphology of this \hii region encouraged us to look for signposts of stellar formation activity at its periphery. In this paper we investigate the stellar formation activity in the three different structures of Sh2-205 and in the large \hi~shell, based on infrared point source catalogues.

We organized the present paper as follows: Sect. 2 gives a brief description of the methods we used to find tracers of stellar formation activity; Sect. 3 shows the main results obtained for the four analyzed regions: \hbox{LBN\,148.11--0.45}, \hbox{SH\,148.83--0.67}, \hbox{SH\,149.25--0.00}, and the \hi~shell. Whether our findings are compatible with the collect and collapse process is discussed in Sect. 4. A summary of the main results is presented in Sect. 5.

\section{Datasets and Selection criteria}

\begin{figure*}
\centering
\includegraphics[angle=0,width=\textwidth]{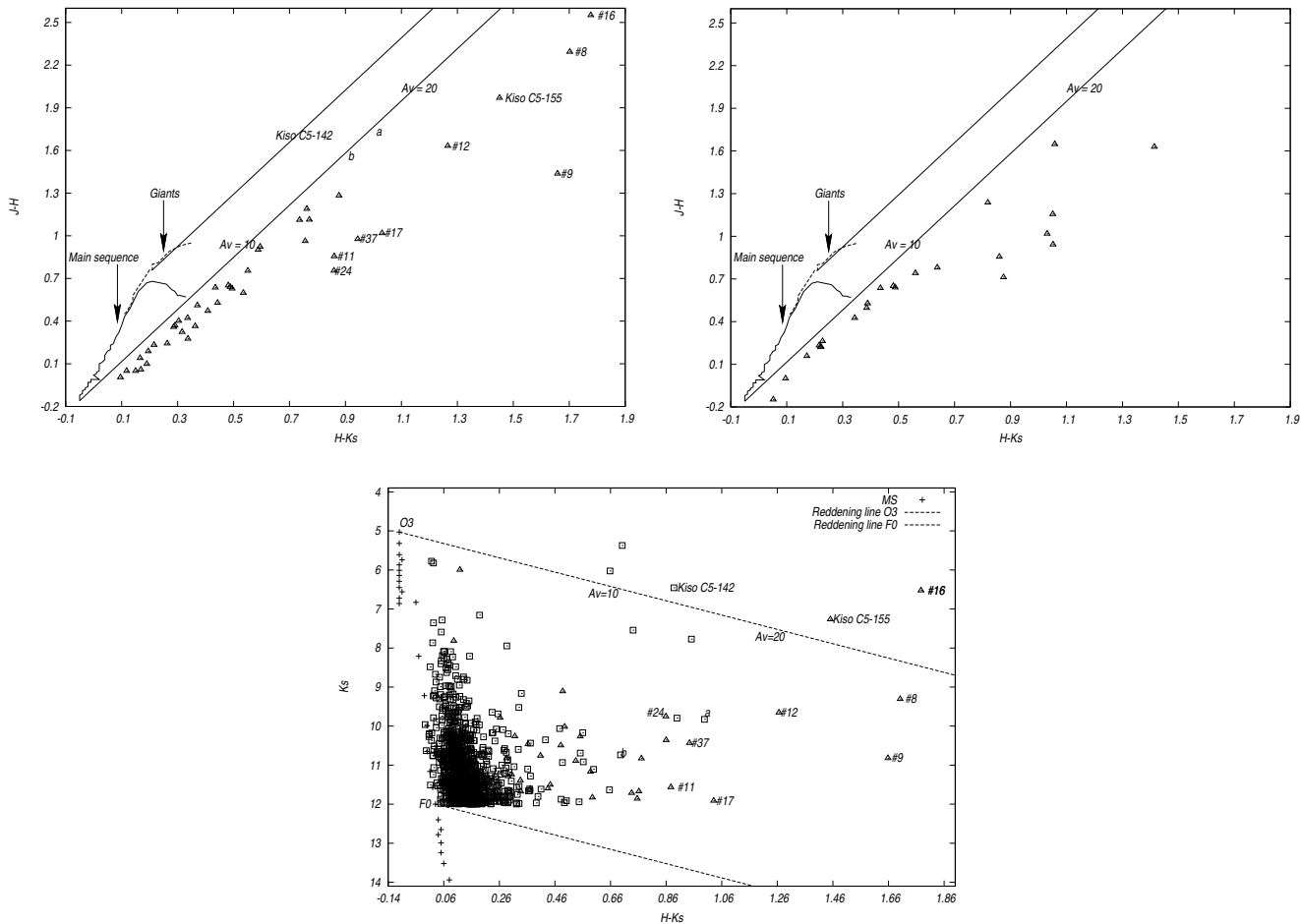}
\caption{\small{CC and CM diagrams of the 2MASS sources towards Sh2-205 and the control area. YSO and main sequence star candidates are indicated by triangles and squares, respectively. {\it Upper panels}: $(H-Ks, J-H)$ diagram of Sh2-205 (left panel) and a control field (right panel). The locations of the derredened early-type main sequence and giant stars are shown. The reddening curves for MO III stars (upper line) and O6-8 V stars (lower line) are indicated. {\it Lower panel}:  $(Ks, H-Ks)$ diagram. The crosses mark the position of the main sequence (MS) without extinction at a distance of 1.0 kpc (Tokunaga 2000, Drilling \& Landolt 2000, Martins \& Plez 2006). The reddening curves for an O3 and a F0 star are shown with dashed lines. Sources with high infrared excess have been identified by the corresponding numbers and letters.}}
\end{figure*}

 In order to find primary indicators of stellar formation activity in the region under study, we used the MSX6C Infrared Point Source Catalogue (Egan \etal2003) in Bands  A (\hbox{8.3 $\,\umu$m} ), C (\hbox{12.1 $\,\umu$m}), D (\hbox{14.7$\,\umu$m}), and E (\hbox{21.3$\,\umu$m}); the 2MASS All-Sky Point Source Catalogue (Cutri \etal2003) in bands $J$ (1.25 $\,\umu$m), $H$ (1.65 $\,\umu$m), and $Ks$ (2.17 $\,\umu$m); and the IRAS Point Source Catalogue \footnote{1986 IRAS catalogue of Point Sources, Version 2.0 (II/125)}. In addition, broadband mid- and far-infrared data supplied by the MSX and IRAS satellites were employed to show the large scale distribution of the dust. Star formation activity was investigated within a box centered at \hbox{{\it (l,b)} = (148\degr 45\arcmin, --0\degr 45\arcmin)} of 2\fdg5~in side.

 $^{12}$CO(1---0) data from Dame, Hartmann \& Thaddeus~(2001) were also used to compare the spatial distribution of the tracers of stellar formation activity with that of the molecular gas. These data have an angular resolution of 8.3 arcmin, a velocity resolution of 1.3 \kms, and an rms noise of 0.05 K.

The criteria used to identify YSO candidates are described in the following paragraphs. 

\subsection{IRAS sources}

A total of 89 IRAS point sources were found projected onto the analyzed region. Junkes, F\"{u}rst \& Reich (1992)'s conditions for young stellar objects are: S$_{100}$ $\ge$ 20 Jy, \hbox{1.2 $\leq$ $\frac{S_{100}}{S_{60}}$ $\leq$ 6.0}, $\frac{S_{100}}{S_{60}}$ $\ge$ 1, and Q$_{60}$+Q$_{100}$ $\ge$ 4, where S$_{\lambda}$ and Q${_\lambda}$ are the flux density and the quality of the IRAS fluxes in each of the observed bands, respectively. Only 7 of the 89 sources can be classified as protostellar candidates following the above mentioned criteria.

\subsection{MSX sources}

The sources were classified based on Lumsden \etal(2002)'s criteria, which allow to select sources taking into account their loci in the (F$_{21}$/F$_{8}$, F$_{14}$/F$_{12}$) diagram. F$_{\lambda}$ denotes the flux in each band. Massive young stellar object (MYSO) candidates have \hbox{F$_{21}$/F$_{8}$ $\ga$ 2} and \hbox{F$_{14}$/$F_{12}$ $\ga$ 1}, while compact \hii regions (CH\,{\sc ii})~present \hbox{F$_{21}$/F$_{8}$ $\ga$ 2} and \hbox{F$_{14}$/$F_{12}$ $<$ 1}. Evolved stars occupy the region \hbox{F$_{21}$/F$_{8}$ $\leq$ 2} and \hbox{F$_{14}$/F$_{12}$ $\leq$ 1}.
 
 A total of 128 MSX sources are projected onto the whole region. Most of the sources are detected only in the highest sensitivity Band A, while 4 per cent of the sources are detected in Band E. For the case of Bands  C and D, 15 per cent of the sources reach the detection limit of the instrument. Of the detected sources, 20 per cent and 33 per cent have reliable fluxes (corresponding to quality flags 3 or 4) in Bands C and D, and in Band E, respectively. Only the sources detected in Bands A and E are included in the present analysis. Among the sources with reliable fluxes, only one can be catalogued as MYSO candidate. Other two sources, which are not detected in Band C, can be classified as young stellar objects, but their identification as MYSO or CHII remain uncertain.

\subsection{2MASS sources}

 A total of 6548 sources were selected from the 2MASS catalogue. These sources have a photometric quality \hbox{{\it Qflg} = AAA}\footnote{JHK Photometric quality of AAA means [jhk]~snr $\ga$10 and [jhk]~cmsig $\la$ 0.10857 and $Ks$ $<$ 12. The condition imposed on Ks allows to reduce source contamination from field stars with spectral types later than F0. This criterium is also compatible with the purpose of this search, which is to identify YSO candidates with high and intermediate masses.} Mean errors are 0.024, 0.03, and 0.023 mag for J, H, and Ks magnitudes, respectively. The nature of the sources can be inferred from their position in the \hbox{($H-Ks$,$Ks$)} CM and \hbox{($H-Ks$,$J-H$)} CC diagrams. Following Comer\'on \& Pasquali (2005), we define the parameter {\it q} as:

\begin{equation} \label{factorq}
q= (J-H) - 1.83 \times (H-Ks)
\end{equation}

This parameter allows identification of sources in different evolutionary stages. Main sequence stars have q-values in the range --0.15 to 0.10, while sources with infrared excess, like YSOs, have q $\la$ --0.15. For giant stars, q $\ga$ 0.10.\\
\indent The selection criteria described above were applied to the Sh2-205 area and to a control field. The last one is centered at {\it (l,b)} = (146\degr 45\arcmin,--0\degr 45\arcmin) with 2\fdg5 $\times$ 2\fdg5 in size. We believe its stellar population is dominated by field stars. Figure 2 shows the CC (upper left panel) and CM (bottom panel) diagrams for main sequence star candidates and sources with infrared excess found towards Sh2-205. The CC diagram of the control field is presented in the upper right panel of Figure 2. The reddening vectors for early type (O6-8 V) and late-type (M0 III) stars (Koornneef 1983) are represented by two parallel lines in the CC diagrams using extinction values from Rieke \& Lebofsky~(1985). The position of the main sequence at a distance of 1.0 kpc is indicated by the crosses in the ($H-Ks$,$Ks$) diagram.\\ 

Towards Sh2-205 we found 41 sources with infrared excess (triangles in Fig. 2), 1082 main sequence stars (represented by squares in the CM diagram of Fig. 2), and 5425 giant stars (not included in the plot). In the CC diagram corresponding to Sh2-205, several sources located just below the reddening curve for an O6-8 V star, having q-values close to 0.1 (not shown in the plot), were included as main sequence star candidates. As regards the control field, we found 23 sources with infrared excess, around half the number identified towards Sh2-205. These sources did not show any spatial clustering. On the contrary, some spatial clustering was found toward Sh2-205 (see Section 3). We note that some of the sources with infrared excess which are situated very close to the reddening vector,
especially at low extinctions, might be early-type stars with small photometric errors, moderate
deviations of the actual reddening law from the adopted one or 
intrinsic scatter in the colors of early-type stars. If this were the case, the diagram woulb be indicating an excess of early-type stars, probably related to the HII region. The condition for Ks-magnitudes causes a region lacking sources in the CM diagram between Ks = 12 and the reddening curve for a F0 star, thus underestimating the number of Herbig Ae/Be candidates in the analyzed region.

\begin{figure*}
\includegraphics[angle=0,width=0.7\textwidth]{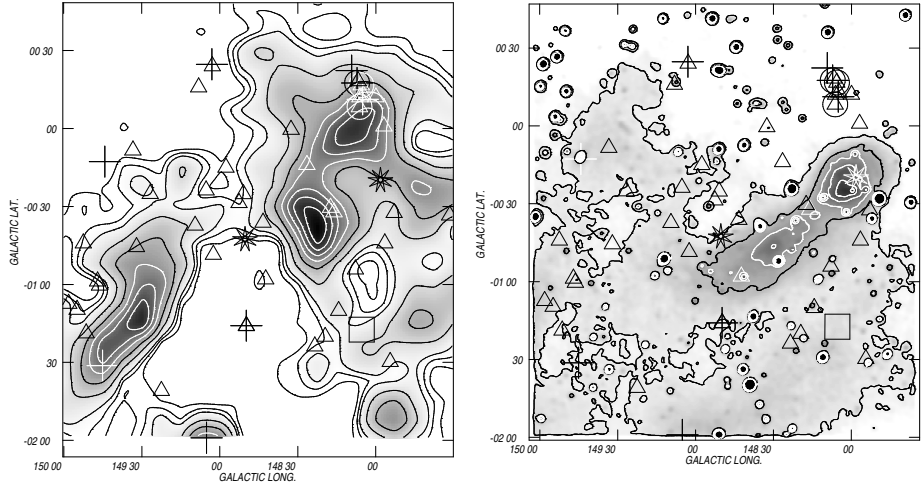}
\caption{\small{{\it Left panel}: YSO candidates superimpossed onto the molecular emission averaged within the velocity range [--0.65, --11.1] \kms.  Contour levels are 0.5, 1.0, 1.5 K, and  from 3 to 15 K in steps of 2 K. {\it Right panel}: YSO candidates overimpossed onto the emission at 1420 MHz. Contour levels are from 5.3 to 6.1 K in steps of 0.2 K. 2MASS, IRAS, and MSX candidates are indicated by triangles, crosses, and circles, respectively. HD\,2431 and HD\,24094 are identified by nine pointed stars. The box indicates the position of NGC\,1444.}}
\end{figure*}

The CC diagram shows a group of four MS star candidates which has the highest values of visual extinction (A$_{v}$ $\approx$ 15 mag). One of them, named Kiso C5-142 in the Simbad database, with values \hbox{($H-Ks$, $J-H$) = (0.89, 1.71)}, and the source with high infrared-excess named Kiso C5-155, with \hbox{($H-Ks$, $J-H$) = (1.47,1.97)}, are actually carbon stars (Alksnis \etal2001). The nature of the other two MS star candidates will be discussed in the following section.

\section{Distribution of YSOs}

In this section, we will examine the spatial distribution of the detected YSO candidates.

Figure 3 shows an overlay of the $^{12}$CO emission distribution in the velocity range [--0.65, --11] \kms~(left panel) and the 1420~MHz image (right panel) and the probable tracers of stellar formation activity. IRAS, MSX, and 2MASS candidates are indicated as crosses, circles, and triangles, respectively. To associate an IR source to a certain \hii region or interstellar bubble within the area under study we considered its proximity to the structure and its correlation with the neutral/ionized associated gas.  We are aware of the fact that no distance information is available in most of the cases. The condition Ks $<$ 12, for the 2MASS sources implies an upper limit to the distance of the sources. The adopted Ks value indicates that we can see an O3 type star with $A_{v} \approx~60~{\rm mag}$ and a B5 type star with $A_{v} \approx~20~{\rm mag}$ at a distance of 1.0 kpc. We discuss our results for the four regions, LBN\,148.11--0.45, \hbox{SH\,148.83--0.67}, \hbox{SH\,149.0--1.30}, and the \hi~shell, separately.

\subsection{Stellar formation in LBN\,148.11--0.45}

\begin{figure*}
\label{fig:sf}
\includegraphics[angle=0,width=0.7\textwidth]{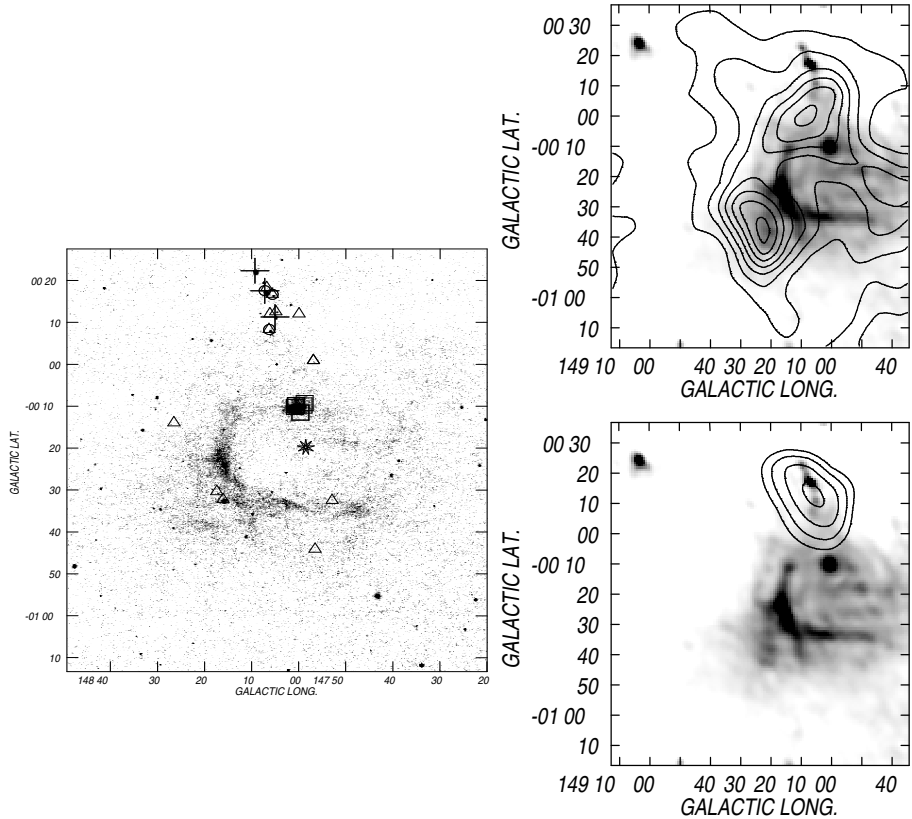}
\caption{\small {\it Left panel}: Image at \hbox{8.3$\,\umu$m}. YSO candidates from 2MASS, IRAS, and MSX sources are indicated by triangles, crosses, and circles, respectively. HD\,24094 is identified by a nine-points star. MS candidates are indicated by squares.{\it Top right panel}: \hbox{100$\,\umu$m} image ({\it grayscale}) and  $^{12}$CO emission distribution ({\it contours}) within the velocity range [--0.65, --11.1] \kms. Contour levels are 1.0, 1.5 K, and from 2.0 to 6.0 K in steps of 1 K.{\it Bottom right panel}: Overlay of the \hbox{100$\,\umu$m} image ({\it grayscale}) and $^{12}$CO emission distribution ({\it contours}) within the velocity range [--31.9, --35.8] \kms. Contour levels are from 1.0 to 2.5 K in steps of 0.5 K.}
\end{figure*}

The distribution of gas and dust in the environs of \hbox{LBN\,148.11--0.45} is shown in Fig. 4. The left panel shows the YSO candidates in the environs of \hbox{LBN\,148.11--0.45} superimpossed onto the emission at 8.3$\,\umu$m. Numerous infrared objects are seen projected onto this region. 2MASS, IRAS, and MSX candidates are indicated by triangles, crosses, and circles, respectively. 

The top right panel displays an overlay of the $^{12}$CO(1--0) emission distribution ({\it contours}) in the range \hbox{[--0.65, --11.1]} \kms~and the infrared emission at \hbox{100$\,\umu$m} ({\it grayscale}), while the bottom right panel depicts an overlay of the $^{12}$CO(1--0) emission distribution ({\it contours}) in the range [--31.9, --35.8] \kms~and the \hbox{100$\,\umu$m} image.

Properties of IRAS, MSX, and 2MASS sources are listed in the Table 1. The designation of the IRAS sources, {\it (l,b)} positions, fluxes at 12$\,\umu$m, 25$\,\umu$m, 60$\,\umu$m, and 100$\,\umu$m, luminosities derived following Yamaguchi \etal (1999), and association with other sources, along with a reference number, are indicated in the table. For the MSX sources we included the candidate designation, the {\it (l,b)} position, the fluxes at 8.3$\,\umu$m, 12.1$\,\umu$m, 14.7$\,\umu$m, and 21.3$\,\umu$m, and the association with other sources. The quality flag of the sources in each band, Q$_{\rm flag}$,  appears indicated in Col. 8 with numbers from 0 to 4. While "0" means that the source has not been detected, "1" to "4" indicate a better detection of the source as the number increases. Identification as MYSO or CHII is included. As regards 2MASS candidates, names, {\it (l,b)} positions, fluxes in the 2MASS IR wavelengths, colours $(J-H)$ and $(H-Ks)$, and association with other sources, are listed.
  
This region can be divided in two areas based on the distribution of the YSO candidates: one close to {\it (l,b)} = (148\degr, +0\degr 15\arcmin) and the other near \hbox{LBN\,148.11--0.45} itself (see Fig. 4).

Twelve candidates are projected near {\it (l,b)} = (148\degr, +0\degr15\arcmin). An inspection of the image at 100$\,\umu$m, shows that the candidates coincide with filamentary emission. This extended feature is not detected at 8.3$\,\umu$m. At this wavelength, only knots of bright emission coincident with the sources are identified.

The YSO candidates are found projected onto the molecular cloud detected in the velocity range \hbox{[--31.9,--35.8] \kms}~(see Fig 3, bottom right panel). The source IRAS\,03523+5343 (IRAS $\#$2) was associated with molecular gas detected in $^{12}$CO at \hbox{v = --34.6 \kms} by Wouterloot \& Brand (1989), who derived a kinematical distance d$_{k}$ $\approx$ 4.2 kpc. Adopting this distance, the infrared luminosity of the source is 6500  L$_{\sun}$ (Yamaguchi \etal 1999). One can wonder whether this star forming region might be an outflow of the \hii\ region LBN\,14811--0.45, and consequently be placed at the same distance. The presence of molecular emission with velocities in the range [--0.65,--11.1] \kms~bordering the ionized region at b --0\degr 10\arcmin\ casts doubts on this interpretation. On the other hand, if the molecular cloud at --34 \kms~were located at the distance of the
\hii\ region, this cloud would have an approaching velocity of at least 25 \kms, 
while we would expect a radial velocity closer than the systemic velocity of the 
\hii\ region for a champagen flow. Moreover, radio emission from 
the ionized gas
in the region between LBN\,14811--0.45 and the star forming area, which
should be present in a champagne flow, is absent.

IRAS\,03529+5345 (IRAS $\#$1) is almost coincident in position with a nebular object present in the NGS-PO Sky Survey and called RNO22 by Cohen (1980), who established from an optical spectrum that it was a F5 star, with relatively high reddening. Based on this
classification and in the apparent magnitude \hbox{m$_{v}$ = 15.3 mag} and visual extinction given by the author, and taking into account \hbox{M$_{v}$ = 3.5 mag} (Drilling \& Landolt 2000), we estimated a distance of \hbox{$\approx$ 500 pc} for the F5 star. 
On the other hand, a chance superposition can not be ruled out. Moreover, the presence of other evidences of active stellar fomation 4\arcmin\ far from IRAS $\#$1 located at larger distances casts doubts on the association of the IRAS source and the F5 star.

As regards IRAS\,03517+5340 (IRAS $\#$3), no additional information is found in the literature. It is also probably related to molecular gas at --33 \kms. If located at 4.2 kpc, its infrared luminosity also suggests a massive object.

MSX sources $\#$4 and $\#$5 and 2MASS sources $\#$7 and $\#$8 are almost coincident with IRAS 
$\#$2, while 2MASS sources $\#$9 and $\#$11 are projected onto the position of IRAS $\#$3. MSX source $\#$6 and 2MASS source $\#$12 are close to the last group. Note that $\#$8, $\#$9, and $\#$12 are among the sources with the highest infrared excess (see Fig. 2). Source $\#$10 is also probably connected to the star forming region at d $\approx$ 4.2 kpc.

As regards, LBN\,148.11--0.45, the 2MASS sources $\#$13 to $\#$18 are placed close to the ionized border. They are projected onto the molecular structure detected within the range \hbox{[--0.65,--11.1]} \kms. Sources $\#$13, $\#$14, $\#$16, and $\#$18 appear projected onto the outer border of the PDR, marked by the emission at 8.3$\,\umu$m (see Fig. 4, left and top right panels).

The 2MASS source $\#$16 has the highest infrared excess in the region [($H-Ks$, $J-H$) = (1.78,2.55)]. It is projected onto the PDR, close to the brightest CO emission region. No additional information on the 2MASS YSO candidates is found in the literature.

 The MSX knot at {\it (l,b)} = (148\degr 0\arcmin,--0\degr 10\arcmin) is an interesting object. As found in Paper I, it is 1.5 arcmin far from the radio continuum point source \hbox{NVSS\, J035327+533601}, which is a non-thermal, probably extragalactic source. Flux densities for this source are 45$\pm$27 mJy and 13$\pm$0.9 mJy at 408 MHz and 1420 MHz, respectively, and  the spectral index is --1.00$\pm$0.2 (Paper I). 
 A total of 5 main sequence candidates are projected onto this knot of emission. They are listed in the bottom part of Table 1. Sources $\#$a and $\#$b have the highest visual extinction among the complete set of main sequence stars. It would be interesting to investigate the true nature of this object.

The difference in distance between the \hii region LBN\,148.11--0.45 and IRAS $\#$2 and $\#$3 suggests that we have identified two independent star forming regions, without apparent physical relation. IRAS $\#$1 is a late type star seen projected onto the environs of \hbox{LBN\,148.11--0.45}, without physical association either with the \hii region or the star forming area at 4.2 kpc.

To sum up, we have identified two star forming regions: one near (l,b) = (148\degr 0\arcmin, +0\degr 15\arcmin) and the other connected to \hbox{LBN\,148.11--0.45}.

As regards the first area, placed at d $\approx$ 4.2 kpc, eleven primary tracers were identified. Six YSO candidates were found in the environs of \hbox{LBN\,148.11--0.45}. Most of them located in the PDR.

\begin{table*}
\label{table:r1}
\begin{center}
\small
\vspace{3mm}
\caption{\small{YSO candidates from the IRAS, MSX, and 2MASS catalogues towards LBN\,148.11--0.45}}
\begin{tabular}{@{}clllrrrrcc}
\hline\hline
\multicolumn{9}{c}{{\bf IRAS sources}} \\
 $\#$ &  Designation &  {\it (l,b)}&     \multicolumn{4}{c}{Fluxes [Jy]}  & $L_{IRAS}$ &Comments\\
             &                         &[$\circ\ \arcmin$,$\circ\ \arcmin$]     &  12$\mu$m  &    25$\mu$m &   60$\mu$m &  100$\mu$m &$[L_{\sun}]$  &\\
\hline
1 & 03529+5345 & 148\degr   9\farcm3 +0\degr  22\farcm32 & 1.4 &1.2     & 13.4 & 40 &18 $\dag$&RNO 22\\
2 & 03523+5343 & 148\degr   7\farcm2  +0\degr  17\farcm52 & 6.1 &30.8 & 87.2 &151 &6500$\dag$ $\dag$ &$\#$4, $\#$5, $\#$7, $\#$8 \\
3 & 03517+5340 & 148\degr  5\farcm04  +0\degr  11\farcm28 & 0.5 &0.5 & 5.7& 31.9 &800$\dag$ $\dag$ &$\#$9, $\#$11\\
\hline
\hline
\multicolumn{9}{c}{{\bf MSX sources}} \\
 \hline
  $\#$ &  Designation  &    {\it (l,b)}&     \multicolumn{4}{c}{Fluxes [Jy]}  & Q$_{\rm flag}$ &Comments\\
        &                           &[$\circ\ \arcmin$,$\circ\ \arcmin$]    &  \hbox{8.3$\,\umu$m}  &  \hbox{12.1 $\,\umu$m} & \hbox{14.7$\,\umu$m} & \hbox{21.3$\,\umu$m} &ACDE  &\\
\hline
\multicolumn{9}{c}{{\bf MYSO candidate}}\\
4 &G148.1201+00.2928 & 148\degr  7\farcm2  +0\degr  17\farcm58 & 3.01 & 6.21 & 8.96 & 21.47 &  4444&$\#$2, $\#$5, $\#$7, $\#$8 \\
 \multicolumn{9}{c}{{\bf Compact \hii region/MYSO candidates}}\\
5 &G148.0930+00.2783 & 148\degr  5\farcm58  +0\degr  16\farcm68 & 0.48 & ... & 0.67 & 2.65 &4013& $\#$2, $\#$4, $\#$7, $\#$8  \\
6&G148.1042+00.1383 & 148\degr  6\farcm4  +0\degr 8\farcm28 & 0.44 & ... & 1.09 & 2.64 & 4011& $\#$12\\
  \hline
\hline
\multicolumn{9}{c}{{\bf 2MASS sources}} \\
\hline
$\#$ & Designation  & {\it (l,b)} [$\circ\ \arcmin$,$\circ\ \arcmin$]&     $J$& $H$ & $Ks$ & $(J-H)$& $(H-Ks)$ &Comments\\
 \hline
 7&03561628+5353025 & 148\degr  6\farcm78  +0\degr  18\farcm3 & 13.62 & 12.43 & 11.67 & 1.19 &0.76 & $\#$2, $\#$4, $\#$5, $\#$8 \\
 8&03560333+5352352 & 148\degr  5\farcm63  +0\degr  16\farcm74 & 13.30 & 11.01 & 9.31 & 2.29& 1.70 & $\#$2, $\#$4, $\#$5, $\#$7\\
 9&03554277+5350074 & 148\degr  4\farcm86  +0\degr   12\farcm54 & 13.92 &  12.48  & 10.82& 1.44&  1.658 &   $\#$3, $\#$11\\
10&03551336+5352390 & 147\degr  59\farcm93  +0\degr  12\farcm06 & 11.63 & 10.97 & 10.49 & 0.65 & 0.48 & \\
11&03554527+5348408 & 148\degr  6\farcm12  +0\degr  12\farcm06 & 13.73 & 12.44 & 11.56 &  1.28 & 0.88 &$\#$3, $\#$9\\
 12&3553004+5345419 & 148\degr  6\farcm30  +0\degr  8\farcm34 & 12.55 & 10.92 & 9.65 & 1.63 &1.27 &$\#$6 \\
13&03540902+5346002 & 147\degr  56\farcm88  +0\degr  0\farcm9 & 10.24 & 9.59 & 9.107 & 0.64&0.49 & \\
 14& 03553869+5315396 &     148\degr  26\farcm52 --0\degr 13\farcm92 & 11.19&  10.83&  10.46&0.37&0.36\\
15&03534163+5308487 & 148\degr  17\farcm39  --0\degr  30\farcm36 & 12.71 & 11.60 & 10.83 & 1.11&0.77 & \\
16&03532733+5308127 & 148\degr  16\farcm07  --0\degr  32\farcm22 & 10.85 & 8.30 & 6.53 & 2.55&1.78 & \\
17&03512640+5322413 & 147\degr  52\farcm92  --0\degr  32\farcm46 & 13.96 & 12.94& 11.91 & 1.02&1.03 & \footnotesize{NVSS\, J035327+533601} \\
 18&03505606+5311217 & 147\degr  56\farcm52  --0\degr  44\farcm1 & 12.67 & 12.03 & 11.59 & 0.64&0.44 & \\
\hline
\multicolumn{9}{c}{{\bf MS candidates towards {\it (l,b)} $\approx$ (148\degr ,--0\degr 10\arcmin)}} \\
a&03533411+5336223&147\degr 58\farcm98 --0\degr 9\farcm84 & 12.56 & 10.83 & 9.83 & 1.73 & 0.99 &\\
b&03534258+5335290&148\degr 0\farcm54 --0\degr 9\farcm72 & 12.72 & 11.43 & 10.74 & 1.29 & 0.69 &\\
c&03533521+5336575&147\degr 58\farcm74 --0\degr 9\farcm24 & 12.25 & 10.69 & 9.79 & 1.56 & 0.89 &\\
d&03534224+5334591&148\degr 0\farcm84 --0\degr 10\farcm14 & 11.57 & 10.78 & 10.35 & 0.79 & 0.43 &\\
e&03533002+5334396 &147\degr 59\farcm64 --0\degr 11\farcm52& 12.71 &12.02& 11.66 &0.69&0.37 &\\
\hline
\hline
\end{tabular}
\end{center}
$\dag$: Calculated adopting a distance of 500 pc\\
$\dag$ $\dag$: Calculated adopting a distance of 4200 pc\\
\end{table*}

\subsection{Stellar formation in SH\,148.83--0.67 and in SH\,149.25--0.00}

Protostellar objects associated with this region are listed in Table 2, which is organized in the same way as Table 1. Figure 3 shows several 2MASS candidates projected onto the radio continuum emission at 1420 MHz. No IRAS and MSX candidates are found in this region.  

Eight YSO candidates were found in this region. Fig. 3 shows the spatial distribution of the candidates onto the atomic neutral gas structure associated with SH\,148.83--0.67 (Paper I). This figure shows the 21--cm HI line emission within the velocity range \hbox {[--25.0,--28.0]} \kms~(Paper I).  Source $\#$26 coincides with the radio continuum point source NVSS~J035347+523208, which  is  probably a pulsar. Flux densities for this source are 33$\pm$16 mJy and 1.94$\pm$0.25 mJy at 408 and 1420 MHz, respectively. The spectral index is --2.27$\pm$0.68 (Paper I).

To summarize, the 2MASS sources are projected onto the HI low emision region and are characterized by low infrared excess compatible with the lack of related molecular and dust emission. This fact casts doubts on the nature of the sources. 2MASS sources $\#$20 and $\#$23 (see Table 2) coincide with HBHA\,5215-03 and HBHA\,5215-01, two emission-line stars (Kohoutek \& Wehmeyer~1999). No additional data about these sources are found in the literature.

\begin{figure}
\includegraphics[angle=0,width=0.50\textwidth]{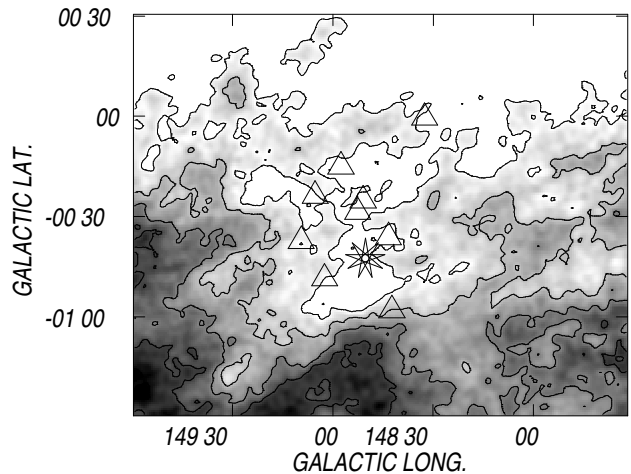}
\caption{\small{YSO candidates overimpossed onto the atomic neutral gas emission distribution associated with SH\,148.83--0.67. The map shows the HI emission distribution in the velocity range from --25.0 to --28.0 \kms. Contour levels are from 60 to 100 K in steps of 10 K. 2MASS candidates are indicated by triangles. HD\,24431 is identified by a nine pointed star.}}
\end{figure}

As regards SH\,149.25--0.00, five 2MASS and one IRAS candidates were found in this region (Table 3). IRAS source \#27 is associated with DSH\,J0402.3+5226. This source was identified as an infrared cluster by Kronberger \etal(2006), reinforcing its identification as a stellar formation tracer. YSO candidates associated with this nebula are listed in Table 3, where the information is organized as in Table 1. 
Additional data  for these sources were not found in the liturature.

In summary, eight candidates were found towards SH\,148.83--0.67, and six ones were identified towards SH\,149.25-0.00. 
The lack of molecular gas linked to these sources strengthens the uncertainty in their true nature.  

\begin{table*}
\centering
\small
\vspace{3mm}
\caption{\small{YSO candidates from 2MASS catalog towards SH\,148.83--0.67.}}
\begin{tabular}{@{}rllcrrrrcc}
\hline
\hline
\multicolumn{9}{c}{{\bf 2MASS sources}} \\
\hline
$\#$ & Designation  &{\it (l,b)} [$\circ\ \arcmin$,$\circ\ \arcmin$]&     $J$& $H$ & $Ks$ & $(J-H)$& $(H-Ks)$ &Comments\\
 \hline
19&03581253+5254467 & 148\degr  57\farcm6 --0\degr  15\farcm0 & 11.51 & 11.41 & 11.22 & 0.09 &0.19 &  \\
20 &03581379+5243109 & 149\degr  5\farcm27  --0\degr  23\farcm76 & 11.64 & 11.17 & 10.76 & 0.47 &0.41 &  HBHA 5215-03\\
21&03565522+5251200 &148\degr 50\farcm94 --0\degr 25\farcm2&  12.03 & 11.43  &10.89&0.59&0.54&\\
22&03564905+5247384&148\degr 52\farcm62 --0\degr 28\farcm62&  12.47&  11.94&  11.50&0.53&0.44&\\
23&03552988+5247463 &148\degr 43\farcm38 --0\degr 36\farcm18&  11.57&  10.82&  10.26&0.75&0.55&HBHA 5215-01\\
24&03573712+5230068 & 149\degr  9\farcm48  --0\degr  37\farcm32 & 11.37 & 10.61 & 9.75 & 0.76 &0.86 & \\
25&03561414+5226030 & 149\degr  2\farcm46 --0\degr 48\farcm54 & 10.89 & 10.58 & 10.26 & 0.32 & 0.32 & \\
26&03535289+5231356 & 148\degr  42\farcm35  --0\degr  58\farcm02 & 12.66 & 11.76 & 11.17 & 0.90 &0.59 & \footnotesize{NVSS~J035347+523208} \\
\hline
\hline
\end{tabular}
\end{table*}

\begin{table*}
\centering
\small
\vspace{3mm}
\caption{\small{YSO candidates from the IRAS and 2MASS
catalogues towards SH\,149.25--0.00.}}
\begin{tabular}{@{}rllrrrrcc}
\hline\hline
\multicolumn{9}{c}{{\bf IRAS sources}} \\
 $\#$ &  Designation & {\it (l,b)}&     \multicolumn{4}{c}{Fluxes [Jy]}  & $L_{IRAS}$ &Comments\\
             &                         &[$\circ\ \arcmin$,$\circ\ \arcmin$]     &  12$\mu$m  &    25$\mu$m &   60$\mu$m &  100$\mu$m &$[L_{\sun}]$  &\\
\hline
27 & 03584+5217 & 149\degr   44.22  --0\degr  12.72 & 1.4& 3.6 & 41.5& 91.1 & 168&DSH\,J0402.3+5226\\
 \hline
\hline
\multicolumn{9}{c}{{\bf 2MASS sources}} \\
\hline
$\#$ & Designation  & {\it (l,b)} [$\circ\ \arcmin$,$\circ\ \arcmin$]&     $J$& $H$ & $Ks$ & $(J-H)$& $(H-Ks)$ &\\
 \hline
28 &04011799+5331341&148\degr 54\farcm72  +0\degr 30\farcm96  &10.06 & 9.88 & 9.7&0.18&0.18&\\
29&04012921+5321001 & 149\degr  2\farcm88  +0\degr  24\farcm12 & 11.94 & 11.54 & 11.24 & 0.40 &0.30 & \\
30&04011993+5311221 & 149\degr  8\farcm16  +0\degr  15\farcm9 & 11.14 & 10.51 & 10.02 & 0.63 & 0.49 & \\
31&03595517+5228084&149\degr 26\farcm78 --0\degr 25\farcm14&  12.31 & 12.12&  11.93&0.19&0.19&\\
32&04014039+5236281 & 149\degr  33\farcm42 --0\degr  8\farcm4 & 12.15 & 11.72 & 11.39 & 0.42 &0.34 & \\
\hline
\hline
\end{tabular}
\end{table*}

\subsection{Stellar formation in the \hi~shell centered at \hbox{{\it (l,b)} = (149\degr 0\arcmin, --1\degr 30\arcmin)}}

\begin{table*}
\centering
\small
\vspace{3mm}
\caption{\small{YSO candidates from 2MASS catalog towards the \hi~shell centered at \hbox{{\it (l,b)} = (149\degr 0\arcmin, --1\degr 30\arcmin)}}}
\begin{tabular}{@{}rllcrrrrcc}
\hline\hline
\multicolumn{9}{c}{{\bf IRAS sources}} \\
 $\#$ &  Designation &  {\it (l,b)}&     \multicolumn{4}{c}{Fluxes [Jy]}  & $L_{IRAS}$ &Comments\\
             &                         &[$\circ\ \arcmin$,$\circ\ \arcmin$]     &  12$\mu$m  &    25$\mu$m &   60$\mu$m &  100$\mu$m &$[L_{\sun}]$  &\\
\hline
33&03494+5204& 148\degr 49\farcm80 --01\degr 15\farcm90&1.1&1.2 &3.9& 20.2&---&CPM\,10, $\#$37\\
34&03530+5117& 149\degr 47\farcm12 --01\degr 31\farcm26&0.8&0.9&10.5&32.5&55&GN\,03.53.0\\
\hline
\hline
\multicolumn{9}{c}{{\bf 2MASS sources}} \\
\hline
$\#$ & Designation  &{\it (l,b)} [$\circ\ \arcmin$,$\circ\ \arcmin$]&     $J$& $H$ & $Ks$ & $(J-H)$& $(H-Ks)$ &Comments\\
 \hline
35&03511239+5256253& 148\degr 07\farcm86  --00\degr 54\farcm18& 11.94 & 11.80  &11.63&0.14&0.17&\\
36&03503912+5239564&148\degr 14\farcm28 --01\degr 10\farcm14&11.16 & 11.11&  10.96&0.05&0.15&\\
37&03531598+5213009 & 148\degr 49\farcm80 --01\degr 15\farcm90 & 12.35&  11.38&  10.43&0.98&0.94&CPM 10, $\#$33\\
38 &03502506+5228548& 148\degr 19\farcm50 --01\degr 19\farcm80&   6.17 & 6.12&  5.99&0.05&0.12&HD\,23800 \\
39 &03502972+5223315&  148\degr 23\farcm46 --01\degr 23\farcm46 & 13.35 & 12.42&  11.83&0.92&0.59&\\
40 &03473627+5237006&  147\degr  54\farcm42 --01\degr 29\farcm76 &12.43  &12.20 & 11.99&0.23&0.22&\\
41&03541653+5132575&   149\degr 22\farcm50 --01\degr 40\farcm92&7.92&  7.91&  7.82&0.001&0.09&HD\,24275\\
42&03585445+5209145 & 149\degr 32\farcm04  --00\degr 45\farcm54 & 11.83&  11.77&  11.59&0.059&0.17&\\
 \hline
\hline
\end{tabular}
\end{table*}

Based on the criteria mentioned in Sect.3, two IRAS and eight 2MASS candidates are projected in the environs of the \hi~and CO shells. They are listed in Table 4.

Source $\#$34 was associated with the reflection nebula GN\,03.53.0 by Magakian~(2003).

IRAS source $\#$33 was identified by Campbell, Persson \& Matthews (1989) as an YSO candidate based 
in its colours ($H-Ks$,$J-H$). Its near-infrared counterpart is 2MASS $\#$37. Its distance is unknown. This source is projected onto the void of molecular emission, suggesting that this object is unconnected to the \hi~shell.

 2MASS sources $\#$38 and $\#$41, which are projected close to the inner border of the molecular ring, were identified as the  near-infrared counterpart of HD\,23800 (B1.5 IVe, Hiltner 1956) and HD\,24275 (A3 V, Rydstrom~1978), respectively. These sources have low infrared excess and lie in the bottom left section of the ($H-Ks$,$J-H$) diagram (Fig. 2), where objects in different evolutionary stages are found (Lada \& Adams 1992).

There is a group of six 2MASS sources with \hbox{{\it l} $\le$ 149\degr 30\arcmin}~(see Fig. 3) which are projected onto the outer border of the molecular ring. These sources may be a connected to Sh2-206, located at \hbox{{\it (l,b)}  = (150\degr 36\farcm18, --00\degr 56\farcm52)}, of 50 arcmin in diameter. Sh2-206 is a blister \hii region catalogued as a star-forming region (Mookerjea \etal 1999). Up to present, neither near nor mid-infrared systematic study was performed to search for  stellar formation tracers in this region. However, the positions of these 2MASS sources suggest that they may be associated with Sh2-206.

To sum up, five YSO candidates probably related to the \hi~and CO shell were identified.

\section{Discussion}

MSX emission at \hbox{21.3$\,\umu$m} is essential to detect embedded sources 
in molecular clouds (e.g Rathborne et al. 2004). However, the least 
sensitivity in  Band E makes it difficult to detect sources at 
\hbox{21.3$\,\umu$m}. This fact leads us to underestimate the number of MSX 
point sources since only sources detected in both Bands A and E have been 
taken into account. Additionaly, only 20-30 per cent of the YSOs can be identified by using 2MASS CC and CM diagrams, as Nielbock, Chini \& M\"uller~(2003) found in the star forming regions OMC 2 and 3. As a consequence, the number of candidates in our analyzed regions is understimated.

Taking into account these constraints, \hbox{LBN\,148.11--0.45} shows a noticeable interesting scenario. The YSO candidates are found in the periphery of the optical nebula, onto or close to the PDR and embedded in the molecular gas. Besides, the age the \hii region (\hbox{4 $\times$ 10$^{6}$ yr}, Paper I) is compatible with the existence of Class I sources (Andr\'e \etal 2000).

This \hii region is evolving in an interstellar medium with an original ambient density of $\approx$ $800~cm^{3}$ (Paper I). These results open the question whether this region is another example of triggered star formation in our Galaxy. The "collect and collapse" model indicates that the expansion of \hii regions on their surroundings creates compressed layers where gas and dust are piled-up between the ionization and the shock fronts. The latter fragmentation of the collected layer ends forming molecular cores where new stars born. The analytical model developed by Whitworth \etal(1994) analyzed the consequences of the dynamical instabilities occurred in the collected layer. They considered three different scenarios where the fragmentation can occur: expanding \hii regions, stellar wind bubbles, and supernova remmants. In all cases, they found that the resulting fragmentation in the shocked layers generates high mass clumps (i.e. $\ge~7~{\rm M}_{\odot}$) which are initially well separated.

For the case of \hii regions, the theory predicts the time at which the fragmentation occurs, $t_{frag}$, the size of the \hii region at that moment, $R_{frag}$, the column density of the shell when the process begins, $N_{frag}$, the mass of the fragments, $M_{frag}$, and their separation along the layers, $r_{frag}$. The parameters required to derive these quantities are the number of Lyman continumm photons emitted per second by the exciting sources, ${\it N}_{Ly}$, the ambient density of the surrounding medium into which the \hii region expands, n$_{0}$, and the isothermal sound speed in the shocked gas, $a_{s}$, which is supposed to be constant.

The analytical expressions are:
\begin{equation}
{\it t}_{frag}~[10^{6}~{\rm yr}]~=~1.56~~a_{.2}^{4/11}~~n_{3}^{-6/11}~~N_{Ly}^{-1/11}
\end{equation}

\begin{equation}
{\it R}_{frag}~[{\rm pc}]~=~5.8~~a_{.2}^{4/11}~~n_{3}^{-6/11}~~N_{Ly}^{1/11}
\end{equation}

\begin{equation}
{\it N}_{frag}~[10^{21}~{\rm cm}^{-2}]~=~6.0~~a_{.2}^{4/11}~~n_{3}^{-6/11}~~N_{Ly}^{1/11}
\end{equation}

\begin{equation}
{\it M}_{frag}~[{\rm M_{\odot}}]~=~23~~a_{.2}^{40/11}~~n_{3}^{-5/11}~~N_{Ly}^{-1/11}
\end{equation}

\begin{equation}
2 {\it r}_{frag}~[{\rm pc}]~=~0.83~~a_{.2}^{18/11}~~n_{3}^{-5/11}~~{\rm N}_{Ly}^{-1/11}
\end{equation}

\noindent where $a_{.2}$ $\equiv$ $\frac{{\it a}_{s}}{0.2~km~s^{-1}}$, $n_{3}$ $\equiv$ $\frac{{\it n}_{0}}{1000~cm^{3}}$

These equations show that $a_{s}$ is an important factor to derive the parameters, while  $n_{0}$ has a lower contribution, and the dependence with $N_{Ly}$ is notoriously weak. A remarkable result from eq. 5 is that the mean mass of the fragments increases as the sound speed grows.

For the case of \hbox{LBN\,148.11--0.45}, the parameters $N_{Ly}$ and $n_{0}$ were derived from the free-free radio continuum emission (Paper I). The rate of Lyman continuum photon emission is at least, \hbox{4~$\times~10^{47}~{\rm s}^{-1}$}, and the ambient density, $800~{\rm cm}^{3}$. Whitworth \etal(1994) pointed out that $a_{s}$ is in the range from 0.2 to 0.6 \kms. For this \hii\ region, the larger $a_{s}$ values lead to unrealistic parameters. For example, for $a_{s}$  = 0.35 \kms, the mass of the fragments increases dramatically ($\approx$ 300 $\rm M_{\odot}$) and their radii reach values larger than the size of the HII region. Acceptable parameters are obtained for the present case by taking into account $a_{s}$ in the range 0.2 to 0.3\kms. In the following, we will adopt 0.2 \kms, which corresponds to typical temperatures of 10 K for the molecular clouds.
 
Considering that the radius of the \hii region is $R_{H II}$ = 7.2 pc, and the dynamical age $t_{dyn}$ \hbox{$=$ $4 \times 10 ^{6}$ yrs}, we compare them with the derived parameters (Table 5). According to the table, $R_{frag}$ $<$ $R_{H II}$ and t$_{dyn}$ $>$ $t_{frag}$, indicating that massive molecular fragments were able to form in the environs of LBN\,148.11--0.45. The distance between fragments is lower than the minimum linear size that the angular resolution of CO data allows to separate (i.e. {\it r} = 2.4 pc at a distance of 1.0 kpc). Better angular resolution data and different line transitions are needed to observe the molecular fragments. Other phenomenon of fragmentation can act simultaneously in this zone. Molecular cores can be formed by radiative-driven collapse, namely the called "Radiation-Driven Implosion" (RDI) theorically developed by Lefloch \& Lazareff (1994). Structures such as cometary globules and elephant trumps can be observational examples of this process
(e.g. Lefloch, Lazareff \& Castets 1997). However, the angular resolutions of optical and infrared observations do not allow to identify the presence of these objects. As a consequence, there is neither evidence nor necessary data to apply the RDI model in this region. Thus, none of the two models can be discarded to explain the stellar formation process in the environs of \hbox{LBN\,148.11--0.45}.

\begin{center}
\begin{table}
\caption{Parameters for the the shocked layers in the environs of LBN\,148.11--0.45.}
\vspace{1.5mm}
\begin{tabular}[h!]{lc}
\hline\hline
Parameter&Value\\
\hline\hline
$M_{frag}$~$[M_{\odot}]$ &33\\
$t_{frag}$~$[10^{6}~yr]$& 2.3 \\
$R_{frag}$~[pc]&5\\
$r_{frag}$~[pc]& 0.6\\
$N_{frag}$~[$10^{21}~cm^{-2}$]&4\\
\hline\hline
\end{tabular}
\end{table}
\end{center}

\section{Summary}

We have analyzed the stellar formation activity in four different regions of Sh2-205.

Based on MSX, 2MASS, and IRAS point source catalogues, we have identified YSO candidates in the region by applying criteria by Junkes, F\"{u}rst \& Reich (1992), Lumsden \etal(2002), and using CM and CC diagrams. Additional information was obtained from Simbad database and broadband mid- and far-infrared images supplied by the MSX and IRAS satellites, and $^{12}$CO(1---0) data.

Our main results can be summarized as follows:

\indent 1.~Six of the seven IRAS, thirty of the fourty-one 2MASS, and the three MSX sources identified as YSO candidates were found projected onto the four structures.

\indent 2.~Of these, eight candidates were found towards  \hbox{SH\,148.83--0.67}, and six candidates were identified towards  \hbox{SH\,149.25-0.00}. However, the lack of molecular gas linked to these sources casts doubts on their nature and their physical association to the structures.

\indent 3.~Five YSO candidates probably related to the \hi~and CO shell centered at \hbox{{\it (l,b)} = (149\degr 0\arcmin, --1\degr 30\arcmin)} were identified. The agent responsible of this structure remains unknown.

\indent 4.~In the environs of \hbox{LBN\,148.11--0.45} we have identified two star forming regions: one near (l,b) = (148\degr 0\arcmin, +0\degr 15\arcmin) and the other connected to \hbox{LBN\,148.11--0.45} itself.

As regards the first area, placed at d $\approx$ 4.2 kpc, eleven primary tracers were identified. 

Six YSO candidates were found in the environs of \hbox{LBN\,148.11--0.45}. Most of them are situated close to the PDR, projected onto molecular material. By applying Whitworth \etal(1994)'s theorical model to this region we find that the formation of massive fragments would be taking place in this region. Higher angular resolution data are necessary to detect the molecular fragments.

5. Finally, two findings can be highlighted: the presence of tracers of star formation activity found in \hbox{LBN\,148.11--0.45} and the requirement of high UV photon flux necessary to keep this \hii region ionized. The combination of these two facts reinforces the argument that HD\,24094 was miss-classified as a B8-type star (Paper I) and the existence of hidden exciting sources responsible for this stellar formation process.

\section{Future work}

Millimetre observations are needed to confirm whether the fragmentation is ocurring in the environs of \hbox{LBN\,148.11--0.45}. Near- and mid-infrared data are important to detect the embedded sources. Spitzer satellite is much more sensitive and has higher spatial resolution than the infrared surveys used in this work (Werner \etal2004). Particularly, IRAC and MIPS 24 $\mu$m colour-colour diagram ([3.6]-[5.8],[8.0]-[24.0]) provides a good observational tool for determing the evolutionary stage of YSOs (Robitaille \etal2006). Unfortunately, no Spitzer data are available for this region. High-quality Spitzer data would allow a more accurate and comprehensive analysis of star formation activity in the environs of LBN\,148.11--0.45.

\section*{Acknowledgments}

We thank the anonymous referee for suggestions that resulted in substantial improvements to the paper. This paper was partially financed by the
Consejo Nacional de Investigaciones Cient\'{\i}ficas y T\'ecnicas
(CONICET)
of Argentina under project PIP 5886/05, Agencia PICT
14018, and UNLP 11/G093. \\
This research has made use of the NASA/ IPAC Infrared Science Archive, which is operated by the Jet Propulsion Laboratory, California Institute of Technology, under contract with the National Aeronautics and Space Administration. This publication makes use of data products from the Two Micron All Sky Survey, which is a joint project of the University of Massachusetts and the Infrared Processing and Analysis Center/California Institute of Technology, funded by the National Aeronautics and Space Administration and the National Science Foundation. The MSX mission is sponsored by the Ballistic Missile Defense Organization (BMDO). This research has made use of the SIMBAD database, operated at CDS, Strasbourg, France.

\label{lastpage}

\end{document}